\documentclass{article}

\usepackage{microtype}
\usepackage{graphicx}
\usepackage{subfigure}
\usepackage{booktabs} 
\usepackage{amsmath}

\usepackage{hyperref}

\usepackage[accepted]{sysml2019}

\sysmltitlerunning{Differentiating SSA-Form Programs}

\begin{document}

\twocolumn[
\sysmltitle{Don't Unroll Adjoint: Differentiating SSA-Form Programs}



\sysmlsetsymbol{equal}{*}

\begin{sysmlauthorlist}
\sysmlauthor{Michael J Innes}{jc}
\end{sysmlauthorlist}

\sysmlaffiliation{jc}{Julia Computing, Inc., London, United Kingdom}

\sysmlcorrespondingauthor{Michael J Innes}{mike.j.innes@gmail.com}

\sysmlkeywords{Machine Learning, SysML}

\vskip 0.3in

\begin{abstract}
This paper presents reverse-mode algorithmic differentiation (AD) based on source code transformation, in particular of the Static Single Assignment (SSA) form used by modern compilers.  The approach can support control flow, nesting, mutation, recursion, data structures, higher-order functions, and other language constructs, and the output is given to an existing compiler to produce highly efficient differentiated code. Our implementation is a new AD tool for the Julia language, called Zygote, which presents high-level dynamic semantics while transparently compiling adjoint code under the hood. We discuss the benefits of this approach to both the usability and performance of AD tools.
\end{abstract}
]



\printAffiliationsAndNotice{}  

\section{Introduction}

Reverse-mode algorithmic differentiation (AD) \cite{speelpenning1980compiling} is at the heart of recent developments in machine learning (ML) and deep learning \cite{baydin2017automatic}. ML systems place extreme demands on the tools used to build them; they typically require the highest performance, yet researchers increasingly need the flexibility of a fully differentiable programming language \cite{innes2018machine}.

AD systems face a tradeoff between providing an expressive, full-featured programming model and producing optimised programs. Current ML frameworks use \textit{tracing} approaches to record the numerical operations in the program, which is simple to implement but requires either constrained semantics or an slow interpreter (Section \ref{practice}). Source-to-source techniques resolve this tradeoff to some extent but have previously been cumbersome or supported only limited semantics (Section \ref{related}).

We present AD over a Static Single Assignment (SSA) representation of programs in a way that supports control flow, higher-order functions and nested derivatives. The differentiated code can be further fed into a traditional compiler such as LLVM \cite{lattner2004llvm}, which results in an extremely efficient derivative program. Further, it opens up the opportunity for robust traditional compiler techniques to be extended to machine learning, enabling kernel fusion or compilation for accelerators with no artificial limitations on the kinds of models that researchers can express.

We additionally introduce Zygote, a working implementation of this technique which augments the Julia compiler \cite{bezanson2017julia} and is designed for use with the Flux machine learning stack \cite{innes:2018}. We discuss Zygote's interaction with Julia's programming model and compiler, and the performance characteristics that result from this combination.

\section{Tapes \& Wengert Lists}

\subsection{Notation \& Background}

Given a target program that outputs a scalar $l$ (typically a loss or objective to be minimised), we write the gradient $\partial l / \partial x$ as $\bar{x}$. For uniformity we do not specify the derivatives of component functions like $\sin(x)$ or $a \times b$ directly in the rules of differentiation, but instead treat these as handled via a higher-order differentiation function $\mathcal{J}$. Given a function $y = f(x_1, x_2, ...)$, we write $y, \mathcal{B}_y = \mathcal{J}(f, x_1, x_2, ...)$; $\mathcal{J}$ returns the usual result $y$ as well as a \textit{pullback function} $\mathcal{B}_y$. Then $\bar{x}_1, \bar{x}_2, ... = \mathcal{B}_y(\bar{y})$; the pullback accepts the gradient with respect to $y$ and returns gradients with respect to each input $x_i$. Pullbacks are linear functions which implement the chain rule for $f$, as in equation \ref{chain}, and for mathematical primitives they are easily written down. Some examples are shown in Table \ref{pullbacks}.

\begin{align}
\label{chain}
    \bar{x} = \frac{\partial l}{\partial x} = \frac{\partial l}{\partial y} \frac{\partial y}{\partial x} = \mathcal{B}_y(\bar{y})
\end{align}

\begin{table}[t]
\caption{Pullbacks for some simple mathematical functions.}
\label{pullbacks}
\vskip 0.15in
\begin{center}
\begin{small}
\begin{sc}
\begin{tabular}{lcccr}
\toprule
Function & Pullback \\
\midrule
$y = a + b$ & $(\bar{y},\bar{y})$ \\
$y = a \times b$ & $(\bar{y} \times b,\bar{y} \times a)$ \\
$y = \sin(x)$ & $\bar{y} \times \cos(x)$ \\
$y = \exp(x)$ & $\bar{y} \times y$ \\
$y = \log(x)$ & $\bar{y}/x$ \\
\bottomrule
\end{tabular}
\end{sc}
\end{small}
\end{center}
\vskip -0.1in
\end{table}

This notation has the benefit of treating program subroutines uniformly with mathematical primitives. In the vector case $\partial y / \partial x$ may be a large Jacobian which we wish to avoid instantiating explicitly. Calling $\mathcal{J}$ with a user-defined $f$ can generate an appropriate pullback via some AD technique (such as the one we describe).

\subsection{Differentiating Wengert Lists}

Consider the following mathematical function, which may be part of our target program. We assume that $y$ is further used to calculate $l$, and that we know $\partial l / \partial y$.
\begin{equation*}
 y = f(a, b) = \frac{a}{a + b^2}
\end{equation*}

We can rewrite this equivalently by naming each intermediate result.
\begin{align*}
 y_1 &= b^2 \\
 y_2 &= a + y_1 \\
 y_3 &= \frac{a}{y_2}
\end{align*}

This form can be viewed as a limited programming language; it is often referred to as a Wengert list, tape or graph \cite{bartholomew2000automatic}. The Wengert list is easy to differentiate. First wrap all function calls with $J$ to create a \textit{primal} version of $f$.
\begin{align*}
 y_1, \mathcal{B}_1 &= \mathcal{J}(\string^, b, 2) \\
 y_2, \mathcal{B}_2 &= \mathcal{J}(+, a, y_1) \\
 y_3, \mathcal{B}_3 &= \mathcal{J}(/, a, y_2)
\end{align*}

Given the gradient $\bar{y}_i$, we can call the pullback $\mathcal{B}_i$ to get gradients for the inputs to $y_i$. Where a variable $x$ is used multiple times, each corresponding pullback produces a contribution to the gradient (the $\bar{a}_i$ below) which must be summed. This is motivated by the multivariable chain rule given in equation \ref{multichain}.
\begin{align}
\label{multichain}
\bar{x} = \frac{\partial l}{\partial x} &= \frac{\partial l}{\partial y_1} \frac{\partial y_1}{\partial x} + \frac{\partial l}{\partial y_2} \frac{\partial y_2}{\partial x} \\
&= \mathcal{B}_{y_1}(\bar{y}_1) + \mathcal{B}_{y_2}(\bar{y}_2)
\end{align}

By applying these steps we can begin with the gradient $\bar{y} = 1$ and proceed in reverse over the list to get $\partial y/\partial a$ and $\partial y/\partial b$. This can be realised either by interpreting the Wengert expression in reverse, or by explicitly creating an adjoint expression as follows.
\begin{align*}
\bar{y}_3 &= 1 \\
\bar{a}_1, \bar{y}_2 &= \mathcal{B}_3(\bar{y}_3) \\
\bar{a}_2, \bar{y}_1 &= \mathcal{B}_2(\bar{y}_2) \\
\bar{a} &= \bar{a}_1 + \bar{a}_2 \\
\bar{b}, \string_ &= \mathcal{B}_1(\bar{y}_1)
\end{align*}

Realising this code as a function, with $\bar{y}_3$ as an argument, creates the pullback for $f$. Inlining all function calls yields an efficient symbolic derivative; the $\mathcal{J}$ notation really is just notation.
\begin{align*}
y_2 &= a + b^2 \\
\bar{y}_2 &= -\frac{a}{y_2^2} \\
y &= \frac{a}{y_2} \\
\bar{a} &= \frac{1}{y_2} + \bar{y}_2 \\
\bar{b} &= 2 b \bar{y}_2
\end{align*}

\subsection{Tapes in Practice}
\label{practice}

To see how Wengert lists can be used to differentiate programs, consider a simple implementation of $x^n$ (for natural $n$).

\begin{verbatim}
function pow(x, n)
  r = 1
  while n > 0
    n -= 1
    r *= x
  end
  return r
end
\end{verbatim}

Typical AD systems use a tracing approach based on operator overloading. The input $x$ is wrapped in a new object which overloads methods such as multiplication ($\times$). $x \times y$ no longer just multiplies $x$ and $y$ but records the operation and its inputs, effectively creating a graph of all basic operations in the program---equivalent to a Wengert list. Invoking \texttt{y = pow(x, 4)} then records the following set of basic operations.
\begin{equation*}
y = ((((1 \times x) \times x) \times x) \times x)
\end{equation*}

The tracing technique is effectively partial evaluation; a language with rich semantics (control flow, data structures, function calls) is heavily specialised on an input to yield a program in a much simpler language (the Wengert list) that can be differentiated.

Simple or not, a program requires evaluation. Tracing AD tools are further split by whether they interpret the trace (``dynamic" frameworks) or compile it (``static" frameworks) \cite{neubig2017dynet}.

Dynamic approaches typically interleave tracing with evaluation of the primal, and have the benefit of preserving the host language's expressive semantics. But they must pay the heavy cost of building and manipulating the graph anew at every iteration, and applying optimisations would cost more time than it saves \cite{paszke2017automatic}. These problems are increasingly important as accelerators become faster than the languages driving them, and optimisations such as operator fusion are needed to get state-of-the-art performance \cite{jiang2018efficient}.

Static systems evaluate the host code only once, record a graph and evaluate it instead of the original program. This comes at a high cost to expressiveness: the graph we recorded for \texttt{pow(x, 4)} above can \textit{only} calculate $x^4$, and if we want richer behaviour we must have mechanisms to insert control flow into the tape. A further fundamental challenge is that traces are an extremely inefficient program representation. The size of the trace for a loop like the above is (size of loop body) $\times$ (number of iterations), leading to a large amount of redundant work for an optimiser; nested loops generate exponentially large traces. Given the infeasibility of running $O(n^2)$ compiler analysis on these graphs, these systems are still interpreted in practice \cite{abadi2016tensorflow}---negating their main theoretical benefit.

These limitations are not fundamental to AD, but instead are limitations of the symbolic form or language that we differentiate---the Wengert list. It would be far more effective to generalise this language, so that it is directly capable of expressing richer programs which can then be fully and efficiently compiled. Happily, just such a generalisation exists via Static Single Assignment (SSA) form.

\section{Static Single Assignment}
\label{ssa}

\subsection{Generalising the Wengert List}

SSA form \cite{cytron1991efficiently} generalises the Wengert list with \texttt{goto}-based control flow, while preserving the explicit data flow that makes analysis straightforward. The primal for the function $f$ above looks as follows in SSA notation, with unique variables labelled $\%1$, $\%2$ and so on.\footnote{For notational convenience we extend SSA with multiple return values, which can be simulated with tuples.}
\begin{align*}
\%1, \%2 &\leftarrow \mathcal{J}(\string^, y, 2) \\
\%3, \%4 &\leftarrow \mathcal{J}(+, x, \%1) \\
\%5, \%6 &\leftarrow \mathcal{J}(/, x, \%3)
\end{align*}

In the adjoint, the important difference from the notation above is the use of underlined references like $\underline{\%6}$, which we refer to as \textit{alphas}. They allow the adjoint code to reuse values from the primal computation without ambiguity, and will be generalised in the case of control flow.
\begin{align*}
\%1, \%2 &\leftarrow \underline{\%6}(1) \\
\%3, \%4 &\leftarrow \underline{\%4}(\%2) \\
\%5 &\leftarrow \%1 + \%3 \\
\%6, \string_ &\leftarrow \underline{\%2}(\%4)
\end{align*}

To see the effect of control flow, consider a branching function.
\begin{equation*}
f(x) = \left\{
\begin{array}{ll}
      x & x > 0 \\
      0.01x & \text{otherwise} \\
\end{array} 
\right. 
\end{equation*}

In SSA form we explicitly test the condition and use a \texttt{goto} to skip the computation of $0.01x$ if it is not necessary.  $\phi$ functions are used to select values from previous blocks; if block $2$ ran then the $\phi$ will return the value of $\%2$, otherwise it just returns $x$ unmodified.\footnote{Though it looks vaguely like BASIC or assembler, the lack of registers or mutable bindings makes SSA closer to a functional representation; basic blocks are equivalent to a set of mutually recursive closures \cite{appel1998ssa}.} The Wengert-list-like code between labels and \texttt{goto} instructions is referred to as a \textit{basic block}.
\begin{align*}
\textbf{block } \#1 \text{:} \\
\%1 &\leftarrow x > 0 \\
&\textbf{goto } \#3 \textbf{ if } \%1 \\
\textbf{block } \#2 \text{:} \\
\%2 &\leftarrow 0.01x \\
\textbf{block } \#3 \text{:} \\
\%3 &\leftarrow \phi (\#1 \rightarrow x, \#2 \rightarrow \%2) \\
&\textbf{return } \%3
\end{align*}

Primal code is created much as before. To construct the adjoint, observe that unrolling the adjoint must be equivalent to constructing the adjoint for an unrolled primal. Thus, all basic blocks must be run in reverse order; there is an (iteration of an) adjoint block for each primal one. To achieve this we invert the primal's control flow graph (CFG) and insert dummy $\phi$ nodes into the primal to record and replay control flow in reverse. (CFG reversal requires there to be a single return node, but multiple returns can easily be merged into one.) After this the basic blocks themselves can be differentiated.

As with the Wengert list, data flow in the adjoint is reversed; a primal SSA definition $\%x$ corresponds to the single usage of the gradient $\overline{\%x}$ with a pullback, and uses of $\%x$ correspond to contributions to the gradient. As SSA definitions dominate their uses, so gradient uses post-dominate their contributions. The complication is that data flow crosses between basic blocks, and a usage of $\%x$ may not actually execute depending on control flow. Thus the adjoint must only take into account gradients that dynamically reach the current block; this can be achieved by propagating gradients in a reversed dataflow analysis of the primal, and inserting zeros and $\phi$ nodes into the adjoint where necessary. For the purpose of finding reaching gradients of $\%x$, primal $\phi$ nodes involving $\%x$ can be treated as equivalent to $\text{identity}(\%x)$.

SSA definitions may take on different values in each iteration of a primal block; alpha nodes refer to the value in the corresponding primal iteration. Given the reversed block order the right semantics can be achieved by storing values on a stack, and alpha nodes are then resolved by popping from the stack \cite{giering1998recipes}. This is not the only possible approach; for example, the values could be recomputed (checkpointing), and mixed approaches are able to make time-space tradeoffs \cite{hascoet2013tapenade}. In a reversible neural network \cite{chang2017reversible}, the core adjoint transformation remains the same but alpha values will be re-calculated in reverse.

The primal thus looks as follows, adding the $\mathcal{J}$ call and dummy $\phi$ node at $\%4$.
\begin{align*}
\textbf{block } \#1 \text{:} \\
\%1 &\leftarrow x > 0 \\
&\textbf{goto } \#3 \textbf{ if } \%1 \\
\textbf{block } \#2 \text{:} \\
\%2, \%3 &\leftarrow \mathcal{J}(\times, 0.01, x) \\
\textbf{block } \#3 \text{:} \\
\%4 &\leftarrow \phi (\#1 \rightarrow false, \#2 \rightarrow true) \\
\%5 &\leftarrow \phi (\#1 \rightarrow x, \#2 \rightarrow \%2) \\
&\textbf{return } \%5
\end{align*}

In the adjoint code we must only apply the pullback $\%3$ to the incoming gradient $\bar{y}$ if block $2$ actually ran. We use $\%4$ to record what control flow happened, and then insert a $\phi$ node to select the correct gradient of $x$.
\begin{align*}
\textbf{block } \#1 \text{:} \\
&\textbf{goto } \#3 \textbf{ if not } \underline{\%4} \\
\textbf{block } \#2 \text{:} \\
\_, \%1 &\leftarrow \underline{\%3}(\bar{y}) \\
&\textbf{goto } \#3 \\
\textbf{block } \#3 \text{:} \\
\%2 &\leftarrow \phi (\#1 \rightarrow \bar{y}, \#2 \rightarrow \%1) \\
&\textbf{return } \%2
\end{align*}

For a more complex example of these rules in practice we take the definition of \texttt{pow} above. The primal code illustrates how loops are represented in SSA form, via $\phi$ nodes. Both relevant variables, $r$ and $n$, are explicitly carried between the two blocks comprising the loop.
\begin{align*}
\textbf{block } \#1 \text{:} \\
\%1 &\leftarrow \phi (\#0 \rightarrow false, \#2 \rightarrow true) \\
\%2 &\leftarrow \phi (\#0 \rightarrow 1, \#2 \rightarrow \%6) \\
\%3 &\leftarrow \phi (\#0 \rightarrow n, \#2 \rightarrow \%5) \\
\%4 &\leftarrow \%3 > 0 \\
&\textbf{goto } \#3 \textbf{ if not } \%4 \\
\textbf{block } \#2 \text{:} \\
\%5 &\leftarrow \%3 - 1 \\
\%6, \%7 &\leftarrow \mathcal{J}(\times, \%2, x) \\
&\textbf{goto } \#1 \\
\textbf{block } \#3 \text{:} \\
&\textbf{return } \%2
\end{align*}

In the adjoint code, we again have two $\phi$ functions in the loop header, effectively tracking $\bar{x}$ ($\%1$) and $\bar{r}$ ($\%2$). Block 1 has two predecessors, block 2 and the implicit block 0 (which corresponds to the return block in the primal). Only $r$ is used in that block (as a return value), so $\bar{x}$ has no gradient contribution and must be initialised to $0$. $x$ is used once in each iteration of the loop, so we accumulate $\bar{x}$ across all iterations.\footnote{Seemingly, so also is $r$. But note each loop iteration sees a \textit{different} definition of $r$, so the gradients are independent. A benefit of SSA form is that this distinction becomes syntactically clear, and need not be handled specially.}
\begin{align*}
\textbf{block } \#1 \text{:} \\
\%1 &\leftarrow \phi (\#0 \rightarrow 0, \#2 \rightarrow \%5) \\
\%2 &\leftarrow \phi (\#0 \rightarrow \bar{y}, \#2 \rightarrow \%3) \\
&\textbf{goto } \#4 \textbf{ if not } \underline{\%1} \\
\textbf{block } \#2 \text{:} \\
\%3, \%4 &\leftarrow \underline{\%7}(\%2) \\
\%5 &\leftarrow \%1 + \%4 \\
&\textbf{goto } \#2 \\
\textbf{block } \#3 \text{:} \\
&\textbf{return } \%2, 0
\end{align*}

\subsection{Handling Language Features}

SSA is a very general representation that does not detail much of a language's semantics (e.g. type system, data structures, memory model). Differentiation depends on these details, largely by way of the primitive definitions provided. For example, the IR may not only contain numerical operations, but also many supporting functions such as for modifying state or manipulating data structures, and we need primitive gradient definitions and pullbacks for these operations.

The most fundamental data structure is the \textit{cons cell}, a tuple of two values like $C = (x_1, x_2)$. If we call \texttt{first(C)} to retrieve the first element we must then find the gradient with respect to $C$ in the adjoint program. We create an \textit{adjoint object} $\bar{C}$, which mirrors the structure of $C$ while storing the gradient of each internal element $(\bar{x}_1, \bar{x}_2)$. Summing adjoint objects sums the elements. The pullbacks for operations on $C$ are as follows.

\begin{center}
\begin{small}
\begin{sc}
\begin{tabular}{lcccr}
\toprule
Function & Pullback \\
\midrule
$C = \text{cons}(x_1, x_2)$ & $(\text{first}(\bar{C}),\text{second}(\bar{C}))$ \\
$y = \text{first}(C)$ & $\text{cons}(\bar{y},0)$ \\
$y = \text{second}(C)$ & $\text{cons}(0,\bar{y})$ \\
\bottomrule
\end{tabular}
\end{sc}
\end{small}
\end{center}

We can now differentiate any function of cons cells.  Any other data structure differs only in number of fields or names of accessor functions.

To handle mutation, consider a one-element ``box" structure $B$. We can $\text{get}(B)$ to retrieve the current stored value, and $\text{set}(B, x)$ to erase that value and replace it with $x$. The adjoint object $\bar{B}$ is also a box, which we retrieve via lookup rather than by pullback return values; a global lookup is necessary to handle the non-local dataflow that mutation introduces. The pullbacks are as follows.

\begin{center}
\begin{small}
\begin{sc}
\begin{tabular}{lcccr}
\toprule
Function & Pullback \\
\midrule
$x = \text{get}(B)$ & $\text{set}(\bar{B}, \text{get}(\bar{B}) + \bar{x})$ \\
$\text{set}(B, x)$ & $(\bar{x} = \text{get}(\bar{B}); \text{set}(\bar{B}, 0); \bar{x})$ \\
\bottomrule
\end{tabular}
\end{sc}
\end{small}
\end{center}

A mutable cons can be seen as a boxed cons or a cons of boxes; in either case it generalises similarly to other mutable data structures. For example, a stack can be implemented as a box containing a cons-based linked list. In general we will want to use more efficient data structures (e.g. stacks in contiguous memory or hash maps), but the box/cons formalism allows us to easily derive appropriate specialised pullbacks for them.

One caveat: pullbacks frequently close over their inputs (for example, both input arrays in matrix multiplication), and if they are mutated the pullback will be incorrect. Arrays must therefore either be immutable, be copied on capture, or have mutations recorded and reversed during the adjoint program. This is generally \textit{not} true for operations on data structures, so things like stacks need no special support.

Given that adjoint code makes use of both stacks and closures, the above ensures that the AD can consume its own output, thus allowing higher-order derivatives via nested application of $\mathcal{J}$ (as in $\mathcal{J}(\mathcal{J},f,x)$).

Closures are just objects with a \texttt{call} method; the fields of the object represent the closure's environment. When calling closures we need to recognise a hidden zeroth argument, the closure itself, and produce an adjoint for that object. In our compiler all functions actually accept this hidden argument---which may be empty as a special case---so both closures and higher-order functions are supported with no extra effort.

These extensions are enough to support a very general subset of the Julia language, thanks to its simple and very uniform semantics. In other cases (such as when class-based objects or lower-level system routines are used), more may be needed. For example, a matrix multiplication might be expressed either by \texttt{A * B} or by \texttt{alloc}/\texttt{free} and passing of pointers, which is harder to differentiate efficiently. For this reason AD is more effective in high-level compiled languages (e.g. Julia, Swift, Rust, Nim) than traditional ones such as C/C++, Fortran and LLVM IR, even though these can all be expressed as SSA.

\section{Optimisation \& Compilation}

\subsection{Interaction with Julia's Compiler}

Aside from correctness, it is important that the adjoint code can be compiled and executed efficiently. A compiler framework must be able to handle the generated code effectively and ultimately produce high-quality machine code. Our implementation in Zygote is designed to interact well with Julia's compiler, and many of the principles are applicable to other languages.

In Zygote, the AD transform is entirely syntactic, and has constraints similar to a Lisp macro (albeit operating with dynamic rather than lexical extent); its compiler interception is similar to previous approaches that extend Julia's compiler at runtime, CUDAnative \cite{besard2018effective} and Cassette \cite{cassette}. Julia's dynamic semantics mean that all function and gradient definitions are (semantically) resolved only at runtime; in general the definition of $f$ and its pullback in $\mathcal{J}(f, x)$ is unknown and could be different each time the code is run. A concrete consequence of this is that we capture pullbacks rather than numerical values directly.

The adjoint code is nevertheless amenable to Julia's standard optimisation heuristics, the most important of which is type inference. Consider the case where the definition of $f$ \textit{can} be inferred statically, as in the $r * x$ in the \texttt{pow} example given above. Since the structure of the pullback is thus also known, we can store just the numerical contents ($r$ and $x$) compactly in memory with no type tags or pointers, and inline the definition of the pullback at its call site. Indeed, if $*$ had instead been $+$, the pullback would be empty, and the compiler could elide the allocation of the stack entirely.

Note that the pullback closure for \texttt{pow} contains (stacks of) pullbacks for the functions it calls, and so on. This can be seen as a kind of tape whose structure defines the adjoint program. However, the stack-based design makes it crucially different from the tapes in other systems: our ``tape" has the structure of the static call graph of the program, \textit{not} the dynamic call graph (as in the traces described in \ref{practice}). This crucial property is what enables Zygote's adjoint code to be effectively statically analysed.

\subsection{Results}

Julia's introspection tools can be used to check that generated output is reasonable. Firstly, we confirm that the code type infers correctly, for example on the adjoint of a simple neural network. This works just as well on larger models such as VGG19, and this level of static analysis is what enables us to target TPUs without tracing \cite{fischer2018automatic}.

\begin{verbatim}
loss(m, x) = sum(m(x))
m = Chain(Dense(10,5,relu),Dense(5,2))
x = rand(10)
@code_typed(gradient(loss, m, x))
# Tuple{NamedTuple{(:layers,),Tuple{
#   Tuple{NamedTuple{(:w, :b, :f),
#     Tuple{Array{Float64,2},
#           Array{Float64,1},
#           Nothing}},
#   NamedTuple{(:w, :b, :f),
#     Tuple{Array{Float64,2},
#           Array{Float64,1},
#           Nothing}}}
# }},Array{Float64,1}}
\end{verbatim}

This type is verbose because it is constructed, by compile-time reflection, as the adjoint of the \texttt{Chain} struct. Since \texttt{Chain} and \texttt{Dense} are functions that happen to have differentiable parameters, this also demonstrates the object-closure relationship described above. Note also that the gradient of \texttt{f}---the activation function of each layer---is statically inferred as non-differentiable; its derivative is always \texttt{nothing}.

After optimisation, the code for \texttt{gradient(pow, 2, 3)} is similar to the following (converted to high-level Julia code for ease of reading).

\begin{verbatim}
function grad_pow(x, n)
  r = 1
  Bs = Tuple{Int,Int}[]
  while n > 0
    push!(Bs, (r, x))
    r *= x
    n -= 1
  end
  dx = 0
  dr = 1
  for i = length(Bs):-1:1
    (r, x) = Bs[i]
    dx += dr*r
    dr = dr*x
  end
  return dx
end
\end{verbatim}

Stacks have low overhead at less than 10 nanoseconds per operation on a typical CPU; this is noticeable compared to scalar numerical operations, but generally negligible in array code. It compares especially favourably to constructing and differentiating a program trace, as in other dynamic AD systems, which has typical overhead in the microseconds per operation \cite{PyTorchYear}.

To confirm this in more realistic cases, Table \ref{benchmarks} provides a set of simple benchmarks between a plain Julia forward pass, Zygote, PyTorch \cite{paszke2017automatic} and ReverseDiff \cite{reversediff} (a tracing-based AD with optional compilation). These mix scalar (\texttt{sincos} and \texttt{loop}) and vector examples to both stress-test AD overhead and show more realistic speedups, respectively.

\begin{table*}[t]
\caption{Benchmarks on some simple functions.}
\label{benchmarks}
\vskip 0.15in
\begin{center}
\begin{small}
\begin{sc}
\begin{tabular}{l|ccccr}
\toprule
Benchmark & Forward & Zygote & PyTorch & ReverseDiff \\
\midrule
SinCos & 15.9ns & 20.7ns & 69,900ns & 670ns \\
Loop & 4.17$\mu$s & 29.5$\mu$s & 17,500$\mu$s & 171$\mu$s \\
LogSumExp & 0.96$\mu$s & 1.26$\mu$s & 219$\mu$s & 15.9$\mu$s \\
Logistic Regression & 4.67$\mu$s & 17.6$\mu$s & 142$\mu$s & 89.9$\mu$s \\
2-Layer MNIST MLP & 27.7$\mu$s & 207$\mu$s & 369$\mu$s & N/A \\
\bottomrule
\end{tabular}
\end{sc}
\end{small}
\end{center}
\vskip -0.1in
\end{table*}

The case without control flow does not even require a stack, and Zygote can match optimised, hand-written gradients in many cases. In cases such as $f(x) = 5x+3$, Julia will type infer the entire call chain, resolve the pullbacks for $*$ and $+$, and inline through all the abstraction (166 different function calls in total) to produce code with only a few integer operations. LLVM then runs constant propagation and produces the following code:

\begin{verbatim}
@code_llvm derivative(x -> 5x+3, 1)

define i64 @"julia_#625_38792"(i64) {
top:
  ret i64 5
}
\end{verbatim}

While LLVM is able to perform powerful optimisations, its knowledge is limited to scalar functions. But there are an increasing number of tensor-aware IRs and compiler stacks \cite{XLA, DBLP:journals/corr/abs-1801-08058, chen2018tvm}, and Zygote's approach to AD makes it much easier to either target these for more advanced optimisations or to apply them on Julia's IR directly---without sacrificing flexibility and abstraction for the researcher.

\section{Related Work}
\label{related}

The most notable existing source-to-source AD systems are Tapenade \cite{hascoet2013tapenade} and Stalin$\nabla$ \cite{pearlmutter2008reverse}. Tapenade is capable of producing very fast code that is amenable to the optimisations of existing Fortran and C compilers. However, it operates directly on source files (requiring ``caller-derives" usage that prevents libraries from abstracting over differentiation), and lacks generality (its output often needs modification before it can be differentiated again). Meanwhile, Stalin$\nabla$ is mathematically general and provides a convenient higher-order-function interface, but only operates on a $\lambda$-calculus IR, a non-standard representation that eschews a large body of work on optimising compilers.

Our contribution is thus to provide a best of both worlds: a system that looks to the user like Stalin$\nabla$, but to the compiler like Tapenade. Our results confirm that we can reach the quality of hand-written derivatives without modifications to an existing optimising compiler.

Myia \cite{van2018automatic} has similar aims in differentiating and compiling a subset of Python. It too generalises Stalin$\nabla$'s $\lambda$-calculus so that closures may have expressions (graphs) as bodies, though does not include mutation or control flow (which is supported by lowering loops to recursion). Differentiating recursion produces a series of nested closures, which at least in principle can be optimised down to a linked list of values; it is then roughly equivalent to Zygote's stacks.

Swift for TensorFlow \cite{swifttf} plans to differentiate a subset of the language using the compiler, with a focus on the error handling and IDE support offered by a static language. The team have discussed the challenges of interfacing with AD in a static type system \cite{swiftad}.

Tangent \cite{wiltschko2017tangent} offers differentiation of a limited subset of Python code, using Python's runtime reflection to retrieve an AST and manipulate it. However, it is mainly aimed at providing intuitive debugging rather than improving performance, since both forwards and backwards passes are still interpreted. Tangent could be generalised by adding a $\mathcal{J}$ operator, though without a compiler this would have the overhead of looking up pullbacks at runtime.

\section{Conclusion}

This paper presents a system for differentiation via the $\mathcal{J}$ function and pullbacks, and uses these to build a system for differentiation via the $\mathcal{J}$ function and pullbacks. Current AD systems which use program-tracing approaches face a fundamental tradeoff between performance and flexibility, but we hope to have shown that this tradeoff is not fundamental. Our new AD, Zygote, supports a full range of language features---from control flow to macros---while producing highly optimised code.

By transforming SSA-form IR we can differentiate rich and expressive programs with extremely low run-time overhead, while opening up opportunities for even more optimisation in future. As SSA is used as an intermediate representation (IR) by many language compilers, differentiation could be added as a first-class language feature to many modern compiled languages, enabling truly differentiable programming.

\section*{Acknowledgements}

Thanks to Avik Sengupta for many insightful conversations. The manuscript was also much improved through feedback from Simon Byrne, Elliot Saba and Deniz Yuret. The implementation was greatly aided by compiler work done for Julia 1.0 by Jarrett Revels, Keno Fischer, Jameson Nash and Jeff Bezanson. The work owes much to enlightening conversations and inspirational work done by people at Julia Computing and the wider Julia community: all of the above along with Viral Shah, James Bradbury, Tim Besard, Simon Danisch, Lyndon White, Shashi Gowda, and many more.


\bibliography{example_paper}

\begin{thebibliography}{29}
\providecommand{\natexlab}[1]{#1}
\providecommand{\url}[1]{\texttt{#1}}
\expandafter\ifx\csname urlstyle\endcsname\relax
  \providecommand{\doi}[1]{doi: #1}\else
  \providecommand{\doi}{doi: \begingroup \urlstyle{rm}\Url}\fi

\bibitem[XLA(2018)]{XLA}
{XLA} overview.
\newblock \url{tensorflow.org/performance/xla}, 2018.
\newblock Accessed: 2018-09-22.

\bibitem[Abadi et~al.(2016)Abadi, Barham, Chen, Chen, Davis, Dean, Devin,
  Ghemawat, Irving, Isard, et~al.]{abadi2016tensorflow}
Abadi, M., Barham, P., Chen, J., Chen, Z., Davis, A., Dean, J., Devin, M.,
  Ghemawat, S., Irving, G., Isard, M., et~al.
\newblock Tensorflow: a system for large-scale machine learning.
\newblock In \emph{OSDI}, volume~16, pp.\  265--283, 2016.

\bibitem[Appel(1998)]{appel1998ssa}
Appel, A.~W.
\newblock {SSA} is functional programming.
\newblock \emph{ACM SIGPLAN Notices}, 33\penalty0 (4):\penalty0 17--20, 1998.

\bibitem[Bartholomew-Biggs et~al.(2000)Bartholomew-Biggs, Brown, Christianson,
  and Dixon]{bartholomew2000automatic}
Bartholomew-Biggs, M., Brown, S., Christianson, B., and Dixon, L.
\newblock Automatic differentiation of algorithms.
\newblock \emph{Journal of Computational and Applied Mathematics}, 124\penalty0
  (1-2):\penalty0 171--190, 2000.

\bibitem[Baydin et~al.(2017)Baydin, Pearlmutter, Radul, and
  Siskind]{baydin2017automatic}
Baydin, A.~G., Pearlmutter, B.~A., Radul, A.~A., and Siskind, J.~M.
\newblock Automatic differentiation in machine learning: a survey.
\newblock \emph{Journal of machine learning research}, 18\penalty0
  (153):\penalty0 1--153, 2017.

\bibitem[Besard et~al.(2018)Besard, Foket, and De~Sutter]{besard2018effective}
Besard, T., Foket, C., and De~Sutter, B.
\newblock Effective extensible programming: Unleashing julia on gpus.
\newblock \emph{IEEE Transactions on Parallel and Distributed Systems}, 2018.

\bibitem[Bezanson et~al.(2017)Bezanson, Edelman, Karpinski, and
  Shah]{bezanson2017julia}
Bezanson, J., Edelman, A., Karpinski, S., and Shah, V.~B.
\newblock Julia: A fresh approach to numerical computing.
\newblock \emph{SIAM review}, 59\penalty0 (1):\penalty0 65--98, 2017.

\bibitem[Chang et~al.(2017)Chang, Meng, Haber, Ruthotto, Begert, and
  Holtham]{chang2017reversible}
Chang, B., Meng, L., Haber, E., Ruthotto, L., Begert, D., and Holtham, E.
\newblock Reversible architectures for arbitrarily deep residual neural
  networks.
\newblock \emph{arXiv preprint arXiv:1709.03698}, 2017.

\bibitem[Chen et~al.(2018)Chen, Moreau, Jiang, Shen, Yan, Wang, Hu, Ceze,
  Guestrin, and Krishnamurthy]{chen2018tvm}
Chen, T., Moreau, T., Jiang, Z., Shen, H., Yan, E., Wang, L., Hu, Y., Ceze, L.,
  Guestrin, C., and Krishnamurthy, A.
\newblock Tvm: end-to-end compilation stack for deep learning.
\newblock In \emph{SysML Conference}, 2018.

\bibitem[Cyphers et~al.(2018)Cyphers, Bansal, Bhiwandiwalla, Bobba, Brookhart,
  Chakraborty, Constable, Convey, Cook, Kanawi, Kimball, Knight, Korovaiko,
  Kumar, Lao, Lishka, Menon, Myers, Narayana, Procter, and
  Webb]{DBLP:journals/corr/abs-1801-08058}
Cyphers, S., Bansal, A.~K., Bhiwandiwalla, A., Bobba, J., Brookhart, M.,
  Chakraborty, A., Constable, W., Convey, C., Cook, L., Kanawi, O., Kimball,
  R., Knight, J., Korovaiko, N., Kumar, V., Lao, Y., Lishka, C.~R., Menon, J.,
  Myers, J., Narayana, S.~A., Procter, A., and Webb, T.~J.
\newblock Intel ngraph: An intermediate representation, compiler, and executor
  for deep learning.
\newblock \emph{CoRR}, abs/1801.08058, 2018.
\newblock URL \url{http://arxiv.org/abs/1801.08058}.

\bibitem[Cytron et~al.(1991)Cytron, Ferrante, Rosen, Wegman, and
  Zadeck]{cytron1991efficiently}
Cytron, R., Ferrante, J., Rosen, B.~K., Wegman, M.~N., and Zadeck, F.~K.
\newblock Efficiently computing static single assignment form and the control
  dependence graph.
\newblock \emph{ACM Transactions on Programming Languages and Systems
  (TOPLAS)}, 13\penalty0 (4):\penalty0 451--490, 1991.

\bibitem[Fischer \& Saba(2018)Fischer and Saba]{fischer2018automatic}
Fischer, K. and Saba, E.
\newblock Automatic full compilation of julia programs and ml models to cloud
  tpus, 2018.

\bibitem[Giering \& Kaminski(1998)Giering and Kaminski]{giering1998recipes}
Giering, R. and Kaminski, T.
\newblock Recipes for adjoint code construction.
\newblock \emph{ACM Transactions on Mathematical Software (TOMS)}, 24\penalty0
  (4):\penalty0 437--474, 1998.

\bibitem[Hascoet \& Pascual(2013)Hascoet and Pascual]{hascoet2013tapenade}
Hascoet, L. and Pascual, V.
\newblock The tapenade automatic differentiation tool: principles, model, and
  specification.
\newblock \emph{ACM Transactions on Mathematical Software (TOMS)}, 39\penalty0
  (3):\penalty0 20, 2013.

\bibitem[Innes(2018)]{innes:2018}
Innes, M.
\newblock Flux: Elegant machine learning with julia.
\newblock \emph{Journal of Open Source Software}, 2018.
\newblock \doi{10.21105/joss.00602}.

\bibitem[Innes et~al.(2018)Innes, Karpinski, Shah, Barber, Stenetorp, Besard,
  Bradbury, Churavy, Danisch, Edelman, et~al.]{innes2018machine}
Innes, M., Karpinski, S., Shah, V., Barber, D., Stenetorp, P., Besard, T.,
  Bradbury, J., Churavy, V., Danisch, S., Edelman, A., et~al.
\newblock On machine learning and programming languages.
\newblock 2018.

\bibitem[Jiang et~al.(2018)Jiang, Chen, and Li]{jiang2018efficient}
Jiang, Z., Chen, T., and Li, M.
\newblock Efficient deep learning inference on edge devices.
\newblock 2018.

\bibitem[Lattner \& Adve(2004)Lattner and Adve]{lattner2004llvm}
Lattner, C. and Adve, V.
\newblock {LLVM}: A compilation framework for lifelong program analysis \&
  transformation.
\newblock In \emph{Proceedings of the international symposium on Code
  generation and optimization: feedback-directed and runtime optimization},
  pp.\ ~75. IEEE Computer Society, 2004.

\bibitem[Neubig et~al.(2017)Neubig, Dyer, Goldberg, Matthews, Ammar,
  Anastasopoulos, Ballesteros, Chiang, Clothiaux, Cohn,
  et~al.]{neubig2017dynet}
Neubig, G., Dyer, C., Goldberg, Y., Matthews, A., Ammar, W., Anastasopoulos,
  A., Ballesteros, M., Chiang, D., Clothiaux, D., Cohn, T., et~al.
\newblock Dynet: The dynamic neural network toolkit.
\newblock \emph{arXiv preprint arXiv:1701.03980}, 2017.

\bibitem[Paszke et~al.(2017)Paszke, Gross, Chintala, Chanan, Yang, DeVito, Lin,
  Desmaison, Antiga, and Lerer]{paszke2017automatic}
Paszke, A., Gross, S., Chintala, S., Chanan, G., Yang, E., DeVito, Z., Lin, Z.,
  Desmaison, A., Antiga, L., and Lerer, A.
\newblock Automatic differentiation in pytorch.
\newblock 2017.

\bibitem[Pearlmutter \& Siskind(2008)Pearlmutter and
  Siskind]{pearlmutter2008reverse}
Pearlmutter, B.~A. and Siskind, J.~M.
\newblock Reverse-mode ad in a functional framework: Lambda the ultimate
  backpropagator.
\newblock \emph{ACM Transactions on Programming Languages and Systems
  (TOPLAS)}, 30\penalty0 (2):\penalty0 7, 2008.

\bibitem[{PyTorch Team}(2018)]{PyTorchYear}
{PyTorch Team}.
\newblock {PyTorch}, a, year in...
\newblock \url{pytorch.org/blog/a-year-in}, 2018.
\newblock Accessed: 2018-09-22.

\bibitem[Revels(2018{\natexlab{a}})]{cassette}
Revels, J.
\newblock Cassette.jl.
\newblock \url{jrevels.github.io/Cassette.jl/v0.1.1}, 2018{\natexlab{a}}.
\newblock Accessed: 2018-09-22.

\bibitem[Revels(2018{\natexlab{b}})]{reversediff}
Revels, J.
\newblock Reversediff.jl.
\newblock \url{github.com/JuliaDiff/ReverseDiff.jl}, 2018{\natexlab{b}}.
\newblock Accessed: 2018-09-22.

\bibitem[Speelpenning(1980)]{speelpenning1980compiling}
Speelpenning, B.
\newblock Compiling fast partial derivatives of functions given by algorithms.
\newblock Technical report, Illinois Univ., Urbana (USA). Dept. of Computer
  Science, 1980.

\bibitem[van Merri{\"e}nboer et~al.(2018)van Merri{\"e}nboer, Breuleux,
  Bergeron, and Lamblin]{van2018automatic}
van Merri{\"e}nboer, B., Breuleux, O., Bergeron, A., and Lamblin, P.
\newblock Automatic differentiation in ml: Where we are and where we should be
  going.
\newblock In \emph{Advances in Neural Information Processing Systems}, pp.\
  8770--8780, 2018.

\bibitem[Wei(2018)]{swiftad}
Wei, R.
\newblock First-class automatic differentition in swift.
\newblock \url{gist.github.com/rxwei/30ba75ce092ab3b0dce4bde1fc2c9f1d}, 2018.
\newblock Accessed: 2018-09-22.

\bibitem[Wei et~al.(2018)Wei, Zheng, et~al.]{swifttf}
Wei, R., Zheng, D., et~al.
\newblock Swift for {TensorFlow}.
\newblock \url{github.com/tensorflow/swift}, 2018.
\newblock Accessed: 2018-09-22.

\bibitem[Wiltschko et~al.(2017)Wiltschko, van Merri{\"e}nboer, and
  Moldovan]{wiltschko2017tangent}
Wiltschko, A.~B., van Merri{\"e}nboer, B., and Moldovan, D.
\newblock Tangent: automatic differentiation using source code transformation
  in python.
\newblock 2017.

\end{thebibliography}
\bibliographystyle{sysml2019}

\end{document}